\documentclass[%
 reprint,
superscriptaddress,
bibnotes,
 amsmath,amssymb,
 aps,
prb,
]{revtex4-1}

\usepackage{graphicx}
\usepackage{dcolumn}
\usepackage{bm}
\usepackage{color}
\definecolor{OliveGreen}{rgb}{0,0.6,0}
\definecolor{auburn}{rgb}{0.43, 0.21, 0.1}
\definecolor{blue_violet}{rgb}{0.54, 0.17, 0.89}
\definecolor{hokie_maroon}{RGB}{99, 0, 49} 
\definecolor{hokie_orange}{RGB}{207, 69, 32} 

\usepackage{appendix}
\usepackage{chngcntr}
\usepackage{etoolbox}
\usepackage{lipsum}
\usepackage{mathtools}

\AtBeginEnvironment{appendices}{%
\addcontentsline{toc}{section}{Appendices}
\counterwithin{figure}{section}
\counterwithin{equation}{section}
\counterwithin{table}{section}
}

\begin{document}


\title{Atomistic Relaxation Process in a Ni$_3$Al Ordered Phase Using Path Probability Method with Vacancy Mechanisms}

\author{Ryo Yamada}
\email{ryamada@imr.tohoku.ac.jp}
\affiliation{Institute for Materials Research, Tohoku University, Sendai 980-8577, Japan}
\author{Tetsuo Mohri}
\email{tmohri@imr.tohoku.ac.jp}
\affiliation{Institute for Materials Research, Tohoku University, Sendai 980-8577, Japan}

\date{\today}

\begin{abstract}
The path probability method (PPM), which is a natural extension of the cluster variation method (CVM) to a time domain, has been employed in a relaxation process of atomic configurations in alloy systems. Although the vacancy mechanism is the main atomic migration process in an alloy system, most studies of PPM have used the spin flipping mechanism (or the direct exchange mechanism) because of the huge computational burden imposed by the vacancy mechanism. In this paper the computational problem is circumvented by treating various path variables in the PPM as cluster probabilities in the CVM, and the vacancy mechanism is explicitly taken into account in the theoretical framework. The method is employed to explore the relaxation process in a Ni$_3$Al ordered phase within the tetrahedron approximation, and the effect of vacancy concentration is investigated.
\end{abstract}

\pacs{Valid PACS appear here}
\maketitle

\section{\label{sec:level1}Introduction}
Recently, first-principles calculation has received much attention because it enables one to gain some insight into various physical properties which are sometimes difficult to elucidate from experiments. The cluster variation method (CVM) \cite{kikuchi1951theory} is one of the most reliable theoretical tools for the derivation of configurational entropy, and has been combined with first-principles band structure calculations to determine the properties of materials at finite temperatures \cite{mohri1993first,terakura1987electronic}. 

Whereas the conventional CVM assumes a rigid lattice of a crystal structure, allowing only a uniform expansion or contraction without changing the symmetry of the lattice, the continuous-displacement cluster variation method (CDCVM) devised by Kikuchi \cite{kikuchi1998space} incorporates local atomic displacements from a rigid lattice by regarding the displaced atoms as being of a different atomic species located in the Bravais lattice point. This conversion of a degree of freedoms from atomic displacements to atomic configurations is a fruitful approach to the exploration of a displacive transition within the realm of replacive transition. This idea of conversion of freedoms has been applied to various internal freedoms, such as the collective displacements of atoms \cite{mohri2013first,kiyokane2018modelling} and spins \cite{yamada2019}, leading to structural and magnetic transitions, respectively.

Although an equilibrium state of an alloy system can be reliably determined by the CVM (or CDCVM), one needs another method in order to obtain kinetic information for a system under transition between equilibrium states. The path probability method (PPM) \cite{kikuchi1966path}, which is a natural extension of the CVM to a time domain, allows one to determine a unique kinetic path without assumptions of near or local equilibrium. The PPM has been applied to order--disorder transitions in binary alloy systems with either a body-centered cubic (bcc) \cite{sato1976kinetics} or face-centered cubic (fcc) \cite{mohri1993kinetic,mohri1997configurational} structure. However, the computational burden is so huge that these applications are still limited to simple model systems. In the calculations for fcc structures, for example, a spin flipping mechanism (or a direct exchange mechanism) is assumed as an atomic migration process, even though a vacancy mechanism is the main mechanism in the atomic diffusion process in alloy systems. The vacancy mechanism was once employed in calculations for bcc alloy systems. However, the pair approximation, which is the simplest approximation in both the CVM and PPM and which demands fewer variables, was used in the special case of equi-atomic alloy composition \cite{sato1976kinetics,gschwend1978kinetics}. In order to employ the vacancy mechanism with a higher order approximation for any alloy composition, it would be necessary to explore an efficient computational approach.

In the present work, the basic idea of the CDCVM, that converts additional freedom of local atomic displacements to those of atomic configurations, is extended to PPM calculations, where path variables in the PPM are treated in the same way as cluster probabilities in the CVM. This conversion of freedoms from kinetic paths to configurational ones offers a concise description of kinetic variables in the vacancy mechanism, and the method paves the way to go beyond a simple pair approximation. We demonstrate a successful example in the calculation of relaxation process in a Ni$_3$Al ordered phase, using the tetrahedron approximation.

The organization of this paper is as follows. In the next section, the theoretical backgrounds of the CVM and PPM are briefly described. In the third section, the calculated results are shown and discussed. The final section is devoted to a conclusion.

\section{\label{sec:level2}Calculation Procedure}

\subsection{\label{sec:level2_1}Cluster Variation Method}
An initial configuration of the PPM calculations is set up by the CVM. The existence of a vacancy to drive diffusion in an A--B binary system requires a CVM formulation for an A--B--$v$ pseudo-ternary alloy system (hereafter, $v$ indicates vacancy). To the best knowledge of the authors, there have been only a few attempts at calculation in A--B--$v$ alloy systems using the CVM. The work of Shinoda et al. \cite{shinoda1992estimation} is one of them, where the equilibrium vacancy concentration in a Ni$_3$Al ordered phase is investigated by regarding it as a Ni--Al--$v$ pseudo-ternary alloy system. Their calculation procedure is followed here, and some essential points are described below. 

A Lennard-Jones type potential is employed to describe the pair interaction energy between two atomic species, $i$ and $j$: 
\begin{equation}
e_{ij}(r)=e_{ij}^0 \left[  \left( \frac{r^0_{ij}}{r} \right)^8 - 2 \left( \frac{r^0_{ij}}{r} \right)^4  \right]  \; , \label{eq:L_J_potential}
\end{equation}
where $e_{ij}^0$ and $r^0_{ij}$ are the fitting parameters, and $r$ is the interatomic distance. The fitting parameters used in reference \cite{shinoda1992estimation} are shown in Table\;\ref{table:L_J_fitting_parameter}, and the same parameters are used here. 

\begin{table}
\begin{center}
\caption{\label{table:L_J_fitting_parameter}The fitting parameters of Lennard-Jones potentials for each pair \cite{shinoda1992estimation}. }
\begin{tabular}{ c  c  c }
$\quad$ & $\quad$ & $\quad$ \\   \hline \hline
$\quad\quad$  &  \quad $e_{ij}^0 \; (\mbox{eV} /\mbox{atom}) $ \quad &  \quad $r_{ij}^0 \; (\mbox{\AA}) $ \quad   \\ \hline
\; Al--Al &  0.5690 & 2.862  \\
\; Ni--Ni & 0.7428 & 2.491  \\
\; Ni--Al &  0.7428 & 2.623  \\ \hline \hline
\end{tabular}
\end{center}
\end{table}

In order to represent the pair interaction energies for metal--vacancy and vacancy--vacancy pairs, the concept suggested by Doyama and Koehler \cite{doyama1976relation} is employed. Doyama and Koehler proposed that the metal--vacancy pair interaction energy can be derived by multiplying the ``mother'' pair energy by 0.35 (this is known as the ``ghost bond model'' \cite{doyama1976relation,shinoda1995cluster}). In the case of a pure metal, A, the mother pair is A--A, so the metal--vacancy and vacancy--vacancy interaction energies become $0.35 \times e_{\mbox{\scriptsize{AA}}}$ and $0.35^2 \times e_{\mbox{\scriptsize{AA}}}$, respectively. On the other hand, when a system consists of two different atomic species, A and B, it is difficult to decide which pair is the mother pair for the A--$v$, B--$v$, and $v$--$v$ pairs. Fortunately, in the case of ordered phases, one can predict the mother pair from the sub-lattices. Since the Ni$_3$Al ordered phase has a L1$_2$ structure, Ni (Al) atoms preferentially exist at the $\beta$ ($\alpha$) sub-lattice. Thus, pair interaction energies for metal--vacancy and vacancy--vacancy pairs can be estimated as shown in Table\;\ref{table:pair_interaction_for_vacancy}. 

\begin{table}
\begin{center}
\caption{\label{table:pair_interaction_for_vacancy}Pair interaction energies for metal--vacancy and vacancy--vacancy pairs at different sets of sub-lattices (where $v$ indicates vacancy) \cite{shinoda1992estimation}. }
\begin{tabular}{ c  c  c  c }
$\quad$ & $\quad$ & $\quad$ & $\quad$ \\   \hline \hline
$\alpha$--$\beta$ & $e_{ij}$ & $\beta$--$\beta$ & $e_{ij}$ \\   \hline 
$\quad$ Al--$v$ $\quad$  & $0.35 \times e_{\mbox{\scriptsize{AlNi}}}$ & $\quad$ Al--$v$ $\quad$ & $0.35 \times e_{\mbox{\scriptsize{AlNi}}}$  \\
Ni--$v$  & $0.35 \times e_{\mbox{\scriptsize{NiNi}}}$ & Ni--$v$ & $0.35 \times e_{\mbox{\scriptsize{NiNi}}}$  \\
$v$--Al  & $0.35 \times e_{\mbox{\scriptsize{AlAl}}}$ & $v$--Al & $0.35 \times e_{\mbox{\scriptsize{NiAl}}}$  \\
$v$--Ni  & $0.35 \times e_{\mbox{\scriptsize{AlNi}}}$ & $v$--Ni & $0.35 \times e_{\mbox{\scriptsize{NiNi}}}$  \\
$v$--$v$  & $0.35^2 \times e_{\mbox{\scriptsize{AlNi}}}$ & $v$--$v$ & $0.35^2 \times e_{\mbox{\scriptsize{NiNi}}}$  \\ \hline \hline
\end{tabular}
\end{center}
\end{table}

The tetrahedron approximation is the minimum meaningful approximation that gives a reasonable accuracy and an acceptable computational burden for an fcc-based alloy system \cite{mohri2017cluster}. Within the tetrahedron approximation, the configurational entropy is given as \cite{kikuchi1974superposition,mohri2013cluster}
\begin{equation}
\begin{split}
S=k_B \; \mbox{ln} \Bigg( \frac{ \left[ \prod_{ij} \left( N \cdot y_{ij}^{\alpha \beta} \right) ! \right]^3 \left[ \prod_{ij} \left( N \cdot y_{ij}^{\beta \beta} \right) ! \right]^3 }{ \left[ \prod_{i} \left( N \cdot x_{i}^{\alpha} \right) ! \right]^{\frac{5}{4}} \left[ \prod_{i} \left( N \cdot x_{i}^{\beta} \right) ! \right]^{\frac{15}{4}} } \;\;\; \\
 \cdot \frac{ N! }{ \left[ \prod_{ijkl} \left( N \cdot w_{ijkl}^{\alpha \beta \beta \beta} \right) ! \right]^{2} } \Bigg) 
\; , \label{eq:entropy_tetrahedron}
\end{split}
\end{equation}
where $N$ is the total number of lattice points, $k_B$ is the Boltzmann constant, and $x_i$, $y_{ij}$, and $w_{ijkl}$ are cluster probabilities of the point, pair, and tetrahedron, respectively. The superscripts in cluster probabilities indicate sub-lattices.

An equilibrium state of a system is determined from the following conditions:
\begin{equation}
\left( \frac{\partial \Omega}{\partial w_{ijkl}^{\alpha \beta \beta \beta}} \right)_{r,T} = 0   \; , \label{eq:free_energy_minimization_cluster_variable}
\end{equation}
\begin{equation}
\left( \frac{\partial \Omega}{\partial r} \right)_{T, x_{i}^{\alpha}, x_{i}^{\beta}, y_{ij}^{\alpha \beta}, y_{ij}^{\beta \beta}, w_{ijkl}^{\alpha \beta \beta \beta}} = 0   \; , \label{eq:free_energy_minimization_distance}
\end{equation}
and 
\begin{equation}
\Omega + \mu_v =0  \; , \label{eq:chemical_potentail_vacancy}
\end{equation}
where $\Omega$ is the grand potential and $\mu_v$ is the effective chemical potential for vacancy. The grand potential, $\Omega$, is given as $\Omega=F+\sum_i \mu_i x_i$, where $F$ is the Helmholtz free energy and $\mu_i$ is the effective chemical potential for atomic species, $i$. The minimization of the grand potential in terms of cluster probability (Eq.\;\eqref{eq:free_energy_minimization_cluster_variable}) is conducted by the natural iteration method (NIM) \cite{kikuchi1974superposition}. Eqs.\;\eqref{eq:free_energy_minimization_cluster_variable} and \eqref{eq:free_energy_minimization_distance} are general conditions for the determination of equilibrium states, but Eq.\;\eqref{eq:chemical_potentail_vacancy} is an additional condition for systems containing vacancies, in which $\mu_v$ is introduced because the free energy should have a minimum in terms of the number of vacancies under the condition that the number of species, A and B (or Al and Ni), is fixed. The detail of the derivation of Eq.\;\eqref{eq:chemical_potentail_vacancy} is not discussed here. Readers interested in the derivation should refer to the original paper \cite{shinoda1992estimation}.

\subsection{\label{sec:level2_2}Path Probability Method}
Whereas the minimization of free energy (or grand potential) is required for a determination of equilibrium states in the CVM, a path probability function (PPF), which is a probability of possible paths in a transition process, is maximized in the PPM to determine the most probable path. Before going into details of the PPF, path variables are defined below, which correspond to cluster probabilities in the CVM. 

Some representative path variables defined in the present work are shown in Table\;\ref{table:path_variables}. It should be noted that the vacancy mechanism is reflected in these path variables, and a direct exchange mechanism, such as A to B (or Al to Ni), is not considered. The basic idea of the CDCVM is used in defining these path variables. In the CDCVM, quasi-lattice points are introduced around each Bravais lattice point and an atom displaced into one of these points is regarded as a different kind of atomic species, which locates at the Bravais lattice point. Hence, the freedom of the atomic displacement is replaced by a configurational freedom of a multicomponent alloy. In the path variables shown in Table\;\ref{table:path_variables}, the path variables for point clusters are treated in the same way as cluster probabilities in the CVM for seven-component systems, suggesting that kinetic freedoms are converted into configurational ones. 

\begin{table}
\begin{center}
\caption{\label{table:path_variables}The path variables defined in this work: the (a) point, (b) pair, and (c) tetrahedron path variables. Some representative path variables are shown for the pair and tetrahedron path variables. Both simplified and conventional expressions for path variables are shown here; e.g., $X_i$ and $\tilde{X}_{i,j}$, respectively. The superscripts of sub-lattices are omitted, and A and B indicate Al and Ni atoms, respectively. }
\includegraphics[scale=0.42]{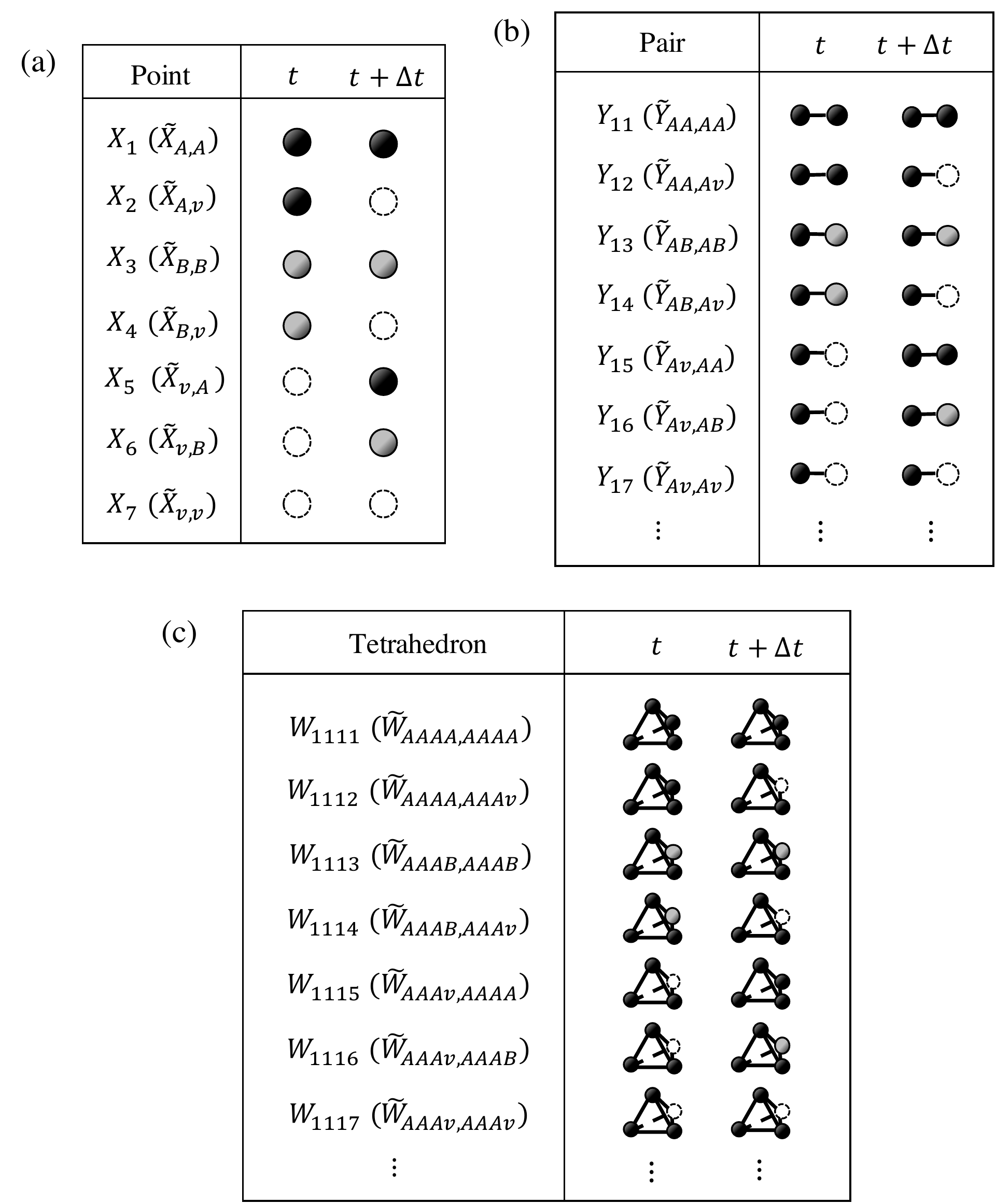}
\end{center}
\end{table}

The conventional expression for path variables is also shown in Table\;\ref{table:path_variables}; e.g., $\tilde{W}_{ijkl,mnop} (t,t+\Delta t)$ describes the transition of tetrahedron configuration $i$-$j$-$k$-$l$ to $m$-$n$-$o$-$p$ in $\Delta t$. Using this conventional expression, the time evolution/devolution of tetrahedron cluster probabilities are given as
\begin{equation}
\begin{split}
w_{ijkl}^{\alpha \beta \beta \beta}(t)=\sum_{mnop} \tilde{W}_{ijkl,mnop}^{\alpha \beta \beta \beta}(t, t+\Delta t) \; , \\
w_{mnop}^{\alpha \beta \beta \beta}(t+\Delta t)=\sum_{ijkl} \tilde{W}_{ijkl,mnop}^{\alpha \beta \beta \beta}(t, t+\Delta t) 
 \; . \label{eq:time_evolution_devolution__tetrahedron_clusters}
\end{split}
\end{equation}
In order to avoid complicacies of notation, a simplified expression for path variables, $X_i$, $Y_{ij}$, and $W_{ijkl}$, is used throughout this study. The path variables satisfy the following normalization and geometrical conditions: 
\begin{equation}
\begin{split}
\sum_i X_i^{\alpha} & = \sum_i X_i^{\beta} = \sum_{ij} Y_{ij}^{\alpha \beta}   \\
&  = \sum_{ij} Y_{ij}^{\beta \beta} = \sum_{ijkl} W_{ijkl}^{\alpha \beta \beta \beta} =1  \;  \label{eq:normalization_condition}
\end{split}
\end{equation}
and
\begin{equation}
\begin{split}
& X_i^{\alpha} =  \sum_{j} Y_{ij}^{\alpha \beta} =\sum_{jkl} W_{ijkl}^{\alpha \beta \beta \beta} \;  \\
X_j^{\beta} &=  \sum_{i} Y_{ij}^{\alpha \beta} =  \sum_{i} Y_{ij}^{\beta \beta} =\sum_{ikl} W_{ijkl}^{\alpha \beta \beta \beta} \; . \label{eq:geometrical_condition}
\end{split}
\end{equation}
The PPF is maximized using NIM as in reference \cite{ohno2005iteration}, where the relation between the cluster probabilities and the path variables, Eq.\;\eqref{eq:time_evolution_devolution__tetrahedron_clusters}, and the normalization condition, Eq.\;\eqref{eq:normalization_condition}, are taken into account. The use of the NIM dramatically reduces the computational burden in the PPM. 

The PPF is defined as the product of three terms: a probability of an atomic jump, $P_1$; a probability of thermal activation for a state change, $P_2$; and the number of equivalent paths, $P_3$. Each term is written as
\begin{equation}
\begin{split}
P_1= \left( \theta \cdot \Delta t \right) ^{\frac{N}{4} \left[ \left( X_2^\alpha + X_4^\alpha + X_5^\alpha + X_6^\alpha \right) + 3 \left( X_2^\beta + X_4^\beta + X_5^\beta + X_6^\beta \right) \right] } \\
\cdot \left( 1- \theta \cdot \Delta t \right) ^{\frac{N}{4} \left[ \left( X_1^\alpha + X_3^\alpha + X_7^\alpha \right) + 3 \left( X_1^\beta + X_3^\beta + X_7^\beta \right) \right] }
  \; ,  \label{eq:P1_term}
\end{split}
\end{equation}
\begin{equation}
P_2=\mbox{exp}\left(  -\frac{\Delta E_{\mbox{\scriptsize{system}}}}{2k_B T} \right)  \; ,  \label{eq:P2_term}
\end{equation}
and
\begin{equation}
\begin{split}
P_3= \frac{ \left[ \prod_{ij} \left( N \cdot Y_{ij}^{\alpha \beta} \right) ! \right]^3 \left[ \prod_{ij} \left( N \cdot Y_{ij}^{\beta \beta} \right) ! \right]^3 }{ \left[ \prod_{i} \left( N \cdot X_{i}^{\alpha} \right) ! \right]^{\frac{5}{4}} \left[ \prod_{i} \left( N \cdot X_{i}^{\beta} \right) ! \right]^{\frac{15}{4}} }  \;\;\; \\
 \cdot \frac{N!}{ \left[ \prod_{ijkl} \left( N \cdot W_{ijkl}^{\alpha \beta \beta \beta} \right) ! \right]^{2} }
\; , \label{eq:P3_term}
\end{split}
\end{equation}
where $\theta$ is the jump probability, $\Delta t$ is the time step, $X_i$, $Y_{ij}$, and $W_{ijkl}$ are path variables of point, pair, and tetrahedron clusters, respectively, $\Delta E_{\mbox{\scriptsize{system}}}$ is a change of internal energy before and after atomic jumps in $\Delta t$, and $T$ is the temperature. The jump frequency, $\theta$, and internal energy change, $\Delta E_{\mbox{\scriptsize{system}}}$, are written as follows:
\begin{equation}
\theta = \theta_0 \cdot \mbox{exp}\left(  -\frac{Q}{k_B T} \right)  \; \label{eq:theta_detail}
\end{equation}
and
\begin{equation}
\begin{split}
\Delta E_{\mbox{\scriptsize{system}}} = 3N \Bigg[ \sum_{ij} e_{ij}^{\alpha \beta} \left( y_{ij}^{\alpha \beta} (t+\Delta t) - y_{ij}^{\alpha \beta} (t)  \right) \;\;\;\;\; \\
+ \sum_{ij} e_{ij}^{\beta \beta} \left( y_{ij}^{\beta \beta} (t+\Delta t) - y_{ij}^{\beta \beta} (t)  \right) \Bigg] \\
+ \frac{N}{4} \Bigg[ \sum_{i} \mu_{i} \left( x_{i}^{\alpha} (t+\Delta t) - x_{i}^{\alpha} (t)  \right) \quad\quad\quad\quad\quad\quad \\
+ 3 \sum_{i} \mu_{i} \left( x_{i}^{\beta} (t+\Delta t) - x_{i}^{\beta} (t)  \right) \Bigg]  \\
 =  3N \left( \sum_{ij} \Delta e_{ij}^{\alpha \beta} Y_{ij}^{\alpha \beta} + \sum_{ij} \Delta e_{ij}^{\beta \beta} Y_{ij}^{\beta \beta}  \right)  \quad\quad\quad\quad\quad  \\
 + \frac{N}{4} \left( \sum_i \Delta \mu_i X_i^\alpha + 3 \sum_i \Delta \mu_i X_i^\beta \right) 
 \; , \label{eq:deltaE_system_detail}
\end{split}
\end{equation}
where $\theta_0$ is an attempt frequency, $Q$ is an activation energy, and $\Delta e_{ij}$ and $\Delta \mu_i$ are the energy changes of pair interaction energies and the chemical potentials in $\Delta t$, respectively. Note that Eq.\;\eqref{eq:entropy_tetrahedron} and Eq.\;\eqref{eq:P3_term} have quite similar forms because Eq.\;\eqref{eq:P3_term} is derived in the same way as the configurational entropy in the CVM. Therefore, Eq.\;\eqref{eq:P3_term} is also called a tetrahedron approximation as well as Eq.\;\eqref{eq:entropy_tetrahedron} in the CVM. 

As mentioned above, the maximization of the PPF is conducted by the NIM as well as the minimization of grand potential in the CVM. Using this approach, a relaxation process of atomic configurations in a Ni$_3$Al ordered phase is examined when it is quenched from 1273\;K to 1073\;K and 1473\;K. These relaxation processes are evaluated by the time dependence of long-range order (LRO) parameters, $\eta$, and their relaxation time, $\tau$, which are, respectively, defined as
\begin{equation}
\eta = \frac{x^\alpha_{\mbox{\scriptsize{Al}}} - x^\beta_{\mbox{\scriptsize{Al}}} + x^\beta_{\mbox{\scriptsize{Ni}}} - x^\alpha_{\mbox{\scriptsize{Ni}}}}{2}  \; \label{eq:long_range_order_definition}
\end{equation}
and
\begin{equation}
\eta (t) = \eta_\infty - (\eta_\infty -\eta_0) \cdot \mbox{exp}\left( -\frac{t}{\tau} \right)  \; , \label{eq:relaxation_time_definition}
\end{equation}
where $\eta_{\infty}$ and $\eta_0$ are, respectively, the LRO parameters at the equilibrium and initial states.

\section{\label{sec:level3}Results and Discussion}
\subsection{\label{sec:level3_1}Equilibrium state}
\begin{figure}
\begin{center}
\includegraphics[scale=0.48]{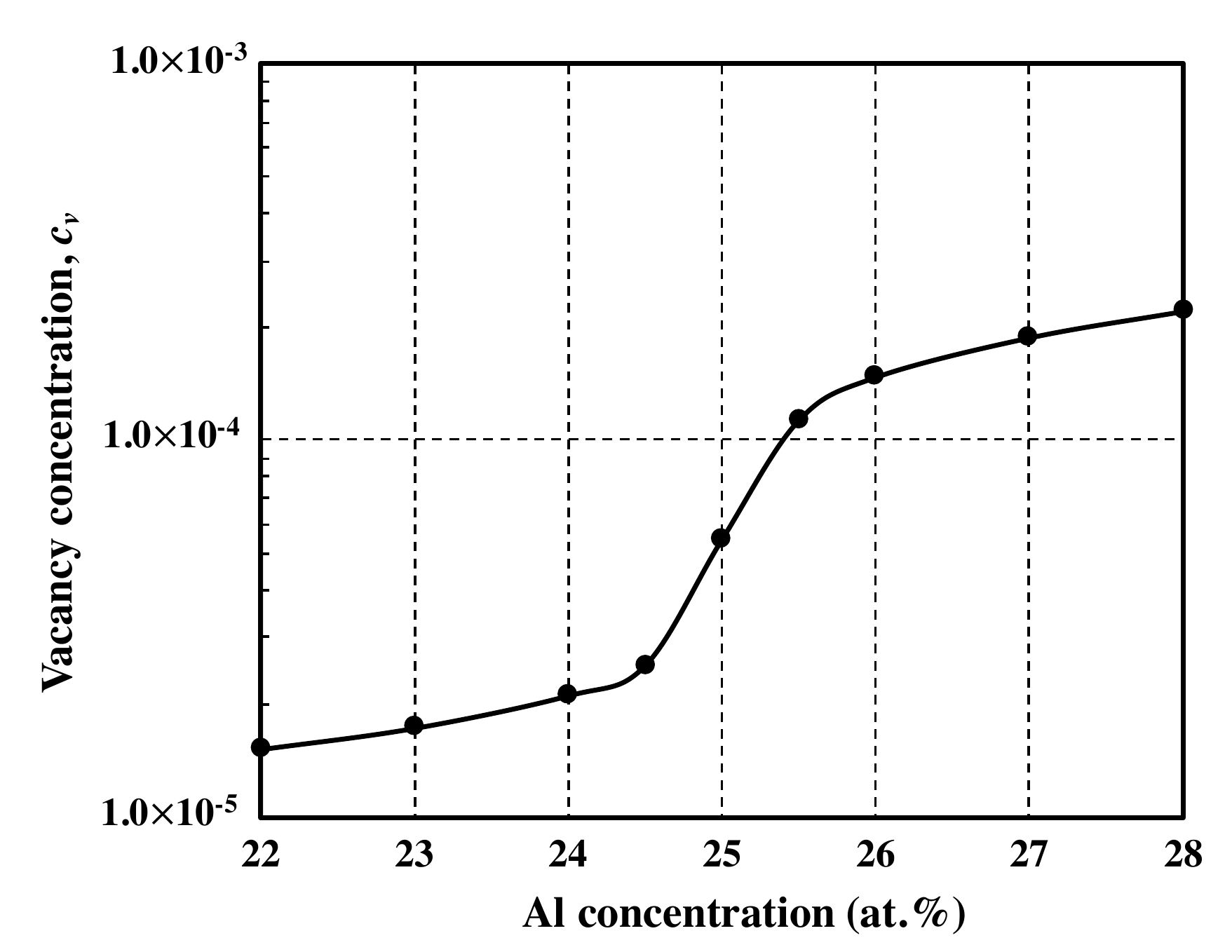}
\caption{\label{fig:vacancy_conc_Al_dependence}The Al concentration dependence of vacancy concentration calculated by CVM at aging temperature, 1273\;K.  }
\end{center}
\end{figure}

The Al concentration dependence of vacancy concentration at 1273\;K calculated by the CVM is shown in Fig.\;\ref{fig:vacancy_conc_Al_dependence}. It shows a step-like dependence and the order of the vacancy concentration is $10^{-5}$. These correspond to the results seen in the preceding work \cite{shinoda1992estimation}. The step-like dependence of vacancy concentration is attributed to the different Al concentration dependences at the two sub-lattices. This interesting behavior strongly depends on the pair interaction parameters used in the calculation (details can be found in reference \cite{shinoda1992estimation}). 

\begin{figure}
\begin{center}
\includegraphics[scale=0.48]{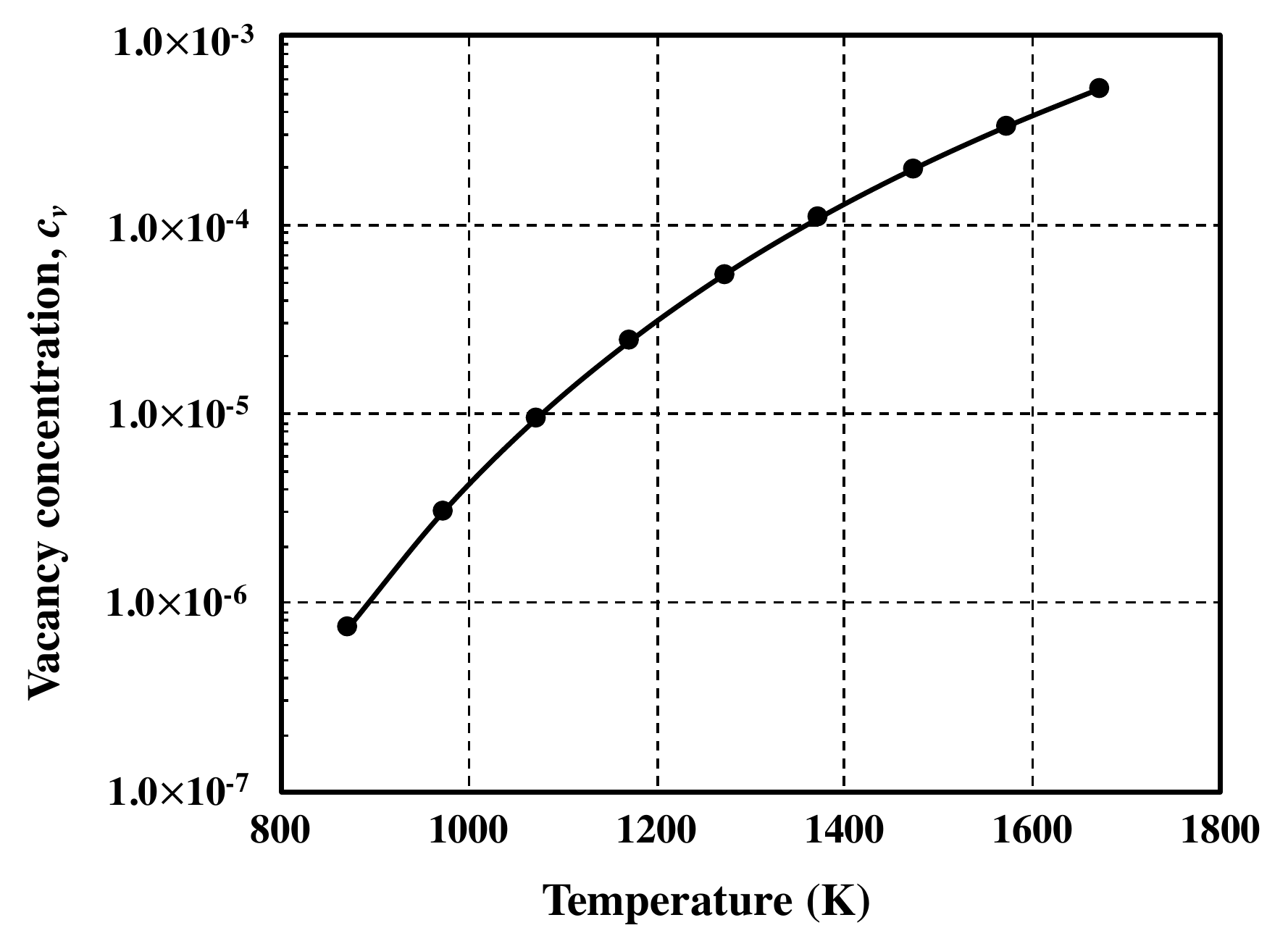}
\caption{\label{fig:vacancy_conc_T_dependence}Temperature dependence of vacancy concentration at 25.0\;at.\%\;Al calculated by CVM. }
\end{center}
\end{figure}

Additionally, the temperature dependence of vacancy concentration at 25.0\;at.\%\;Al is shown in Fig.\;\ref{fig:vacancy_conc_T_dependence}. It can be seen that the vacancy concentration increases with temperature. This is due to the contribution of configurational entropy at high temperatures. These atomic configurations (or cluster probabilities) determined from the CVM are used as an initial state in the PPM calculations (Sec.\;\ref{sec:level3_2}). 

\subsection{\label{sec:level3_2}Kinetic path to the equilibrium state}
\begin{figure}
\begin{center}
\includegraphics[scale=0.55]{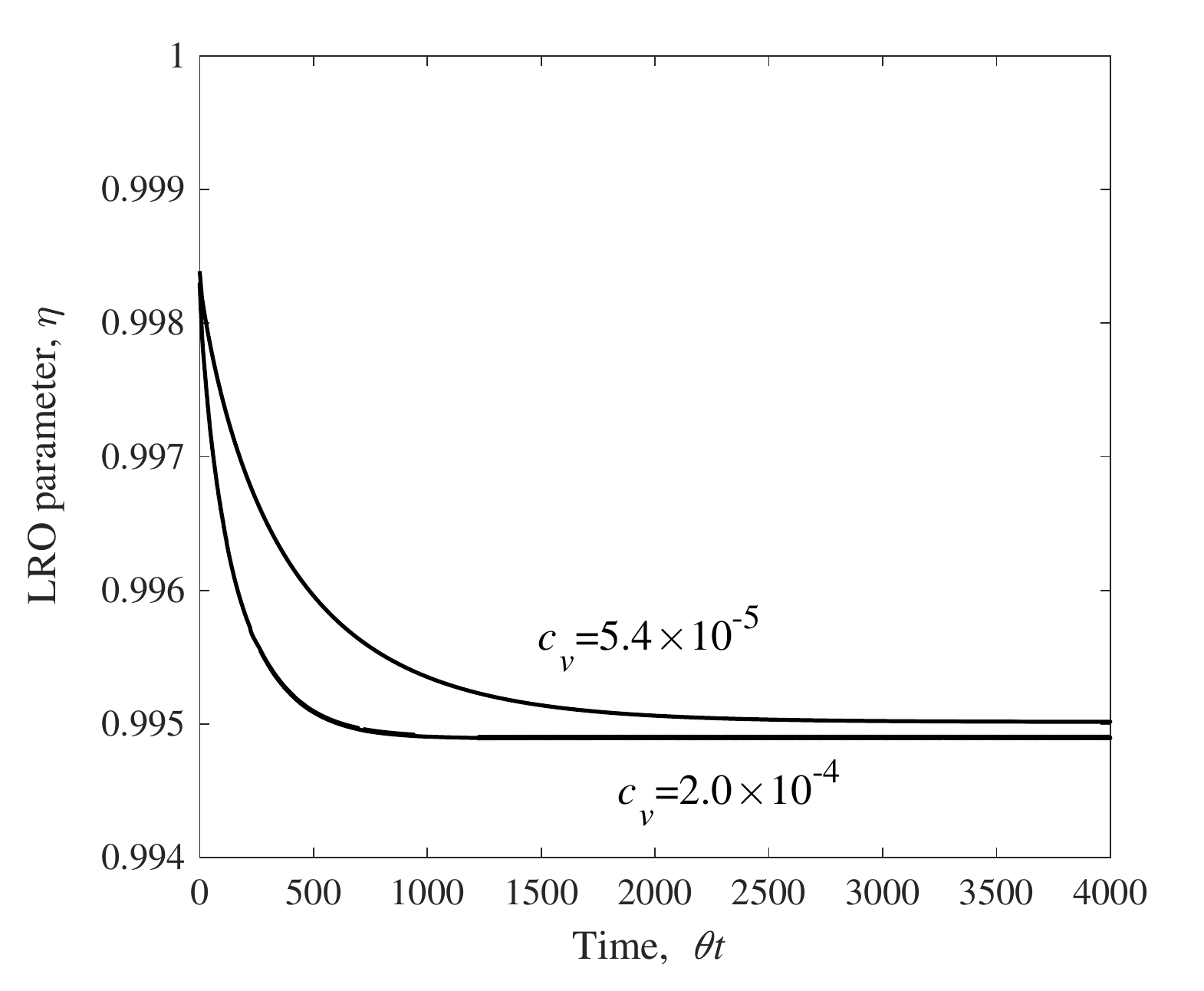}
\caption{\label{fig:LRO_time_dependence_1473}Time evolutions of LRO parameter at 25.0\;at.\%\;Al when the system is quenched from 1273\;K to 1473\;K. The horizontal line is normalized by the jump probability, $\theta$.  }
\end{center}
\end{figure}

\begin{figure}
\begin{center}
\includegraphics[scale=0.55]{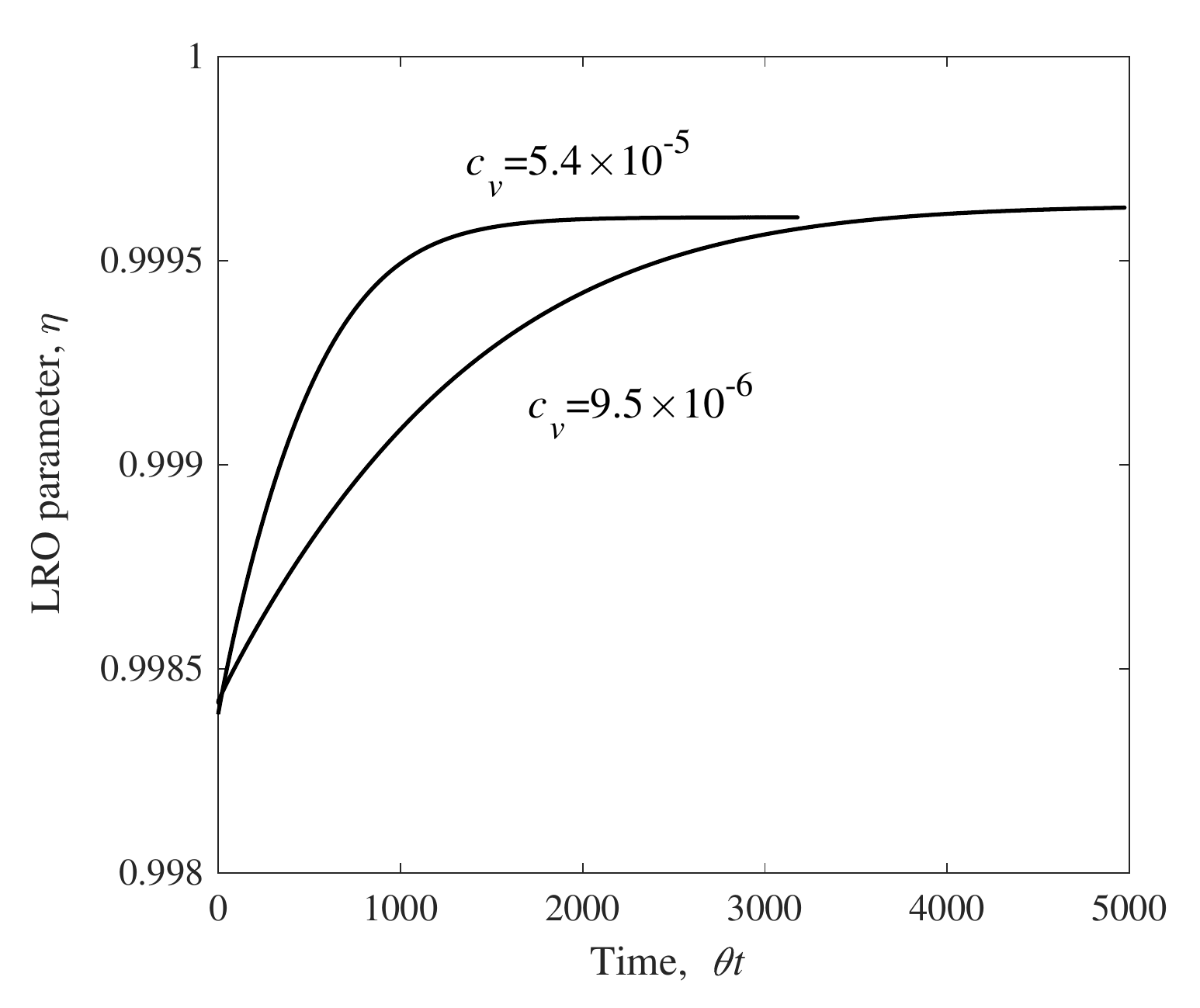}
\caption{\label{fig:LRO_time_dependence_1073}Time evolutions of LRO parameter at 25.0\;at.\%\;Al when the system is quenched from 1273\;K to 1073\;K. The horizontal line is normalized by the jump probability, $\theta$. }
\end{center}
\end{figure}

The time evolution of LRO parameters, $\eta$, at 25.0\;at.\%\;Al during the isothermal aging process at 1473\;K and 1073\;K are shown in Figs.\;\ref{fig:LRO_time_dependence_1473} and \ref{fig:LRO_time_dependence_1073}, respectively. The horizontal lines are normalized by the jump probability, $\theta$. For these calculations the concentration of vacancy is fixed at the equilibrium value of an initial state (i.e., $c_v$ at 1273\;K) or at the value at the isothermal aging temperature (i.e., $c_v$ at 1473\;K or 1073\;K). Figures\;\ref{fig:LRO_time_dependence_1473} and \ref{fig:LRO_time_dependence_1073} show that the LRO parameters become smaller (larger) with time for the aging at 1473\;K (1073\;K). The long time limits of their values are confirmed to be the equilibrium ones independently calculated for each temperature by the CVM (see Sec.\;\ref{sec:level3_1}).

From the results of the LRO parameter, the relaxation time, $\tau$, is obtained by fitting into Eq.\;\eqref{eq:relaxation_time_definition}. The relaxation times derived from aging at 1473\;K and 1073\;K are shown in Table\;\ref{table:relaxation_time_result}. From these results, it is confirmed that the more (fewer) vacancies exist, the smaller (larger) the relaxation time the system has. If the vacancy concentrations are allowed to alter during aging, it is expected that the relaxation time would be at some value between the values obtained here. 

\begin{table}
\begin{center}
\caption{\label{table:relaxation_time_result}Relaxation times derived from Figs.\;\ref{fig:LRO_time_dependence_1473} and \ref{fig:LRO_time_dependence_1073}. These relaxation times are normalized by the jump probability, $\theta$. }
\begin{tabular}{ c  c  c  c }
$\quad$ & $\quad$ & $\quad$ & $\quad$ \\   \hline \hline
\; & \; & $c_v$ & \; \\   \hline
$\quad$ & $\quad$ $9.5 \times 10^{-6}$ & $\quad$ $5.4 \times 10^{-5}$ & $\quad$ $2.0 \times 10^{-4}$ \\  \hline 
1473\;K  & -- & 391.4 & 158.8  \\
1073\;K  & 1170.0 & 458.8 & --  \\ \hline \hline
\end{tabular}
\end{center}
\end{table}

Note that the horizontal lines in Figs.\;\ref{fig:LRO_time_dependence_1473} and \ref{fig:LRO_time_dependence_1073} are normalized by the jump frequency, $\theta$. The real time dependence of LRO parameters are determined if the attempt frequency, $\theta_0$, and the activation energy, $Q$, are known. These values would be calculated from a band structure calculation, but this is beyond the scope of the present study.

\section{\label{sec:level4}Conclusions}
An idea of conversion of freedoms used in CDCVM is applied to the PPM framework, where kinetic freedoms are converted into configurational freedoms, and the relaxation process in a Ni$_3$Al ordered phase is explored by taking into account the vacancy mechanism as an atomic migration process. It is confirmed that a higher (lower) vacancy concentration in the system leads to a faster (slower) relaxation time. 

It is noted that in the present work the timescale of the calculated results is normalized by the jump probability. The jump probability is related to the activation energy and attempt frequency. The incorporation of realistic values can be achieved by combining it with electronic structure calculations, which will be reported elsewhere.

\section*{ACKNOWLEDGEMENT}
This study is partly supported by a project (P16010) commissioned by the New Energy and Industrial Technology Development Organization (NEDO), and by the Structural Materials for Innovation of the Cross ministerial Strategic Innovation Promotion Program (SIP) of Japan Science and Technology (JST). One of the authors (TM) appreciate their supports. Also, we express sincere appreciation to Dr. Y. Yamada at Materials and Energy Division, CSRC, Beijing for his stimulating discussions.

\bibliographystyle{ieeetr}
\bibliography{ref}

\end{document}